# An AC susceptometer for the characterization of large, bulk superconducting samples

**P Laurent[1], J F Fagnard[1], B Vanderheyden[1], N Hari Babu[2], D A Cardwell[2], M Ausloos[3] and P Vanderbemden[1]**

(1) SUPRATECS and Department of Electrical Engineering and Computer Science B28, University of Liège, Sart-Tilman, B-4000 Liège, Belgium

(2) IRC in Superconductivity, University of Cambridge, Madingley Road, Cambridge CB3 0HE, United Kingdom

(3) SUPRATECS and Department of Physics B5, University of Liège, Sart-Tilman, B-4000 Liège, Belgium

E-mail : laurent@montefiore.ulg.ac.be

**Abstract.** The main purpose of this work was to design, develop and construct a simple, low-cost AC susceptometer to measure large, bulk superconducting samples (up to 32 mm in diameter) in the temperature range 78-120 K. The design incorporates a double heating system that enables a high heating rate (25 K/hour) while maintaining a small temperature gradient (< 0.2 K) across the sample. The apparatus can be calibrated precisely using a copper coil connected in series with the primary coil. The system has been used successfully to measure the temperature dependence of the AC magnetic properties of entire RE-Ba-Cu-O [(RE)BCO] bulk superconducting domains. A typical AC susceptibility measurement run from 78 K to 95 K takes about 2 hours, with excellent temperature resolution (temperature step ~ 4 mK) around the critical temperature, in particular.

# 1. Introduction

The measurement of the AC magnetic susceptibility as a function of temperature is a fast, contactless and powerful tool for studying magnetic or superconducting materials [1-4]. In the case of superconductors, the real ($\chi'$) and imaginary parts ($\chi''$) of the AC susceptibility represent measures of flux exclusion due to induced shielding currents and magnetic losses due to the movement of flux lines, respectively [1-8]. The AC susceptibilty measurement technique became particularly popular following the discovery of High-$T_c$ Superconductors (HTS) in 1986 and many research laboratories now invariably possess either commercial or home-built AC susceptometer systems. Despite the popularity of this measurement technique, AC susceptometer systems (both commercial and home-made) for measurements at cryogenic temperatures [9-14] generally accommodate samples of relatively small size (typically < 10 mm diameter). The relatively recent development of large, single grain bulk superconductors by melt processing techniques, however, has generated a need to characterize samples magnetically over larger diameters and volumes [15-18]. Such samples, which are usually of disc or cylindrical geometry, have a typical size of 30 mm diameter and 10 mm thickness [19-20] and are being used increasingly in numerous permanent magnet-like engineering applications such as energy storage flywheels, magnetic bearings, motors and generators due to their ability to trap large magnetic fields [19]. The present work describes a simple and easy-to-use experimental arrangement suitable for measuring precisely the AC magnetic properties of large samples as a function of temperature. In particular, we discuss an efficient calibration method and how the problems associated with the temperature non-uniformity within the sample can be overcome.

The target application is the study of bulk superconductors belonging to the RE-Ba-Cu-O [(RE)BCO] family of materials (RE = rare earth Y, Nd, Sm, Eu, etc.) fabricated by melt processing techniques such as top seeded melt growth (TSMG) [21]. Within this system, Gd, Sm and Nd-based materials have recently attracted much attention [20,22] due to their relatively high superconducting transition temperatures ($T_c$) (e.g. 95 K for NdBCO compared to 92 K for well-established YBCO) [20]. An increase of $T_c$ of only 3 K in these materials gives rise to substantially improved superconducting properties at liquid nitrogen temperatures (77 K). An undesirable feature, however, is that these materials may suffer from an inhomogeneous distribution in $T_c$ all over the sample volume [22]. In addition, regions of low $T_c$ can also occur in large single grains due to in-homogeneity in oxygen content. It is essential, therefore, to determine accurately the transition temperature of the bulk single domain after the synthesis process. For example, a (RE)BCO single grain sample exhibiting a broad superconducting transition width may indicate that the sample requires further annealing in $O_2$ or that a further heat treatment in Ar is needed to improve the $T_c$ distribution [22]. The standard magnetic characterisation technique for bulk superconductors involves mapping the trapped flux over of the top surface with a miniature Hall probe [23]. This measurement, however, is usually performed at a fixed temperature (typically 77 K) and, as a result, does not yield any clear information about magnetic phenomena at the superconducting transition temperature. To date, the standard practice has been to cut the single domain into small specimens and to characterise their magnetic properties as a function of temperature individually, using AC susceptometry or DC magnetometry (SQUID, VSM) [24,25]. The procedure is time-consuming and destructive, which is particularly inconvenient given that large bulk materials may take up to 2 weeks to process. The availability of an AC susceptometer designed specifically to accommodate large samples (up to 32 mm in diameter in the present work) is therefore essential for the characterization of as-processed, bulk single grain superconductors material, in addition to studying the magnetic behaviour (vortex dynamics and AC losses) of these materials under variable magnetic field. Such a system can also potentially detect microstructural defects in the sample, such as cracks and grain boundaries, which are likely to impede the superconducting current flow and modify the AC magnetic signal.

The practical requirements for the experimental arrangement are dictated by the properties of the materials that are to be characterized. Firstly, the system has to exhibit good temperature resolution in the 88-95 K temperature range (i.e. the vicinity of the superconducting transition), which, in this case, can be achieved using liquid nitrogen as a coolant. Secondly, the total characterization time should be as short as possible, e.g. typically less than 3-4 hours. Thirdly, the sample temperature should be as homogeneous as possible throughout the measurement. A reasonable requirement is that the heating rate used for the measurement should not induce a temperature difference of greater than 0.2 K within the sample. These three characteristics can be satisfied simultaneously and easily in the case of commercial (small sample) susceptometers or those described in the literature [1-4, 9-14]. For samples of larger size, however, achieving such a combination of properties represents a significant technical challenge due the large thermal time constants associated with large samples. In order to allow for a reasonable measurement time, temperature stabilization step-by-step using a temperature controller is not appropriate; the system has to work in a temperature sweep mode. Therefore, particular care has to be taken in heating the sample during measurement. It is also of interest to be able to modulate the heating rate in order to increase the temperature resolution in the vicinity of the superconducting transition temperature. Finally, the large sample susceptometer should be able to superimpose a DC magnetic field to an AC magnetic field in order to measure field-dependent AC properties of large, single grains.

In this paper we report the experimental design of the susceptometer, the measurement technique, the calibration method, the heating system and the cryogenic aspects of the design. We report and discuss the performance of the heating system and present illustrative measurements on large samples of bulk (RE)BCO superconductors fabricated by top seeded melt growth (TSMG).

## 2. Experiment

*2.1. Coil assembly*

A susceptometer consists generally of a magnetizing (primary) coil and two pick-up (secondary) coils wound electrically in series opposition. The electro-motive force (emf) induced across ideal pick-up coils (i.e. wound with the same number of turns and in identical geometries) is zero in the absence of a sample. Inserting a magnetic sample centred in one of the secondary coils results in the generation of a non-zero signal that is directly proportional to the AC susceptibility, $\chi = \chi' - j\chi''$ [1-4], of the specimen. A fundamental constraint of the coil-assembly is that the magnetizing field should be as homogeneous as possible over the entire sample volume, which is typically difficult to achieve for large samples. In our system, the primary (magnetizing) coil consists of two quasi-identical separate coils (labeled "1r" and "1s" in Fig. 1) placed electrically in series. Each primary coil contains its own secondary coil ("2r" and "2s") and the sample is inserted into one of the secondary coils (2s). The experimental configuration adopted here has two advantages. Firstly, the sample can be located precisely at the centre of one of the (long) primary coils, i.e. in the zone where the magnetic field is most uniform. This is particularly significant for characterizing large samples. Secondly, the primary coil containing the sample (1s) can be inserted easily in another (larger) coil used to superimpose a DC magnetic field (0.5T in the present case), as shown in Fig. 1(a). Note that the two primary coils are placed well apart (0.5 m) in order to avoid inductive coupling.

The geometrical details of the primary and secondary coils are given in Table 1. The homogeneity of the magnetic field produced by the primary coil was checked by numerical modelling using Quickfield® software. The field uniformity within a cylindrical region (height: 12 mm: diameter: 30 mm) centred in the primary coil was found from this model to be better than 1 %. In practice, it is impossible to wind two identical sets of coils, which results in a slight offset voltage across the pick-up coils in the absence of a sample [1-3]. Several methods can be used to reduce this effect. These include either (i) moving the sample from one pick-up coil to the other and the subtracting the signals [2-4], or (ii) using passive or active offset adjustment elements [11,26-28]. Achieving a successful compensation process can be difficult due to the variation of the balance point with temperature, applied field and frequency. The present design attempts to minimize the offset voltage without using compensation. The two pairs of pick-up coils were wound with great care, which resulted in a difference in their electrical characteristics (e.g. resistance, inductance) of less than 1 %. In addition, the offset signal was tuned finely by moving slightly the second pick-up coil along the axis of its primary coil. This procedure resulted in an offset signal of typically less than 0.5 % of the nominal voltage appearing across the pick-up coils in the presence of a perfectly diamagnetic sample (i.e. $\chi' = -1$, $\chi'' = 0$).

Each primary coil was wound on an ultra high molecular weight polyethylene (PE-UHMW) cylinder of diameter 50 mm diameter, which was found to have excellent mechanical properties from room temperature down to 77 K. Each secondary coil was wound on a 40 mm diameter cylinder placed within the primary coil, as shown in Fig. 1(b). The coils were stabilized mechanically with GE varnish. The sample was placed in a vacuum vessel hollowed out of the support of one pick-up coil, as illustrated in Fig. 1(b). The whole system was inserted in a liquid nitrogen bath at T = 77 K.

Using the coil parameters shown in Table 1, the primary coil generates a primary field $H_a$ = 19.1 A/m ($\mu_0 H_a$ = 24 µT) for a 1 mA drive current. The current is supplied by a Keithley 6221 AC current source and the primary coils can carry a maximum constant current of 6 A at 77 K. The signal across the pick-up coil is read by a lock-in amplifier (Perkin Elmer 5210) which acts a phase-sensitive voltmeter [29]. The reference signal is the voltage drop across a resistor in series with the primary coils.

*2.2. Calibration and demagnetizing field considerations*

The susceptometer described here is designed to accommodate cylindrical samples of small aspect ratio (i.e. of typical length/diameter ~ 1:3). The measured signal, therefore, needs to be corrected for the effects of demagnetization fields; i.e. one has to distinguish between the (true) internal susceptibility $\chi_{int}$ (= d$M$/d$H$) and the measured, or "external", susceptibility $\chi_{ext}$ (= d$M$/d$H_a$), where $M$, $H$ and $H_a$ denote the sample magnetization, the internal field and the applied field, respectively [3]. These quantities are related by;

$$\chi_{int} = \frac{\chi_{ext}}{1 - D\chi_{ext}}, \qquad (1)$$

where $D$ is the demagnetization factor [3]. In the particular case of ellipsoids of revolution, the demagnetizing field is uniform inside the material and the values of $D$ can be calculated analytically [30]. In general, however, the demagnetization field is not uniform; the value of the demagnetizing field should be computed numerically and an average effective value of $D$ should be considered [31,32]. Average values of $D$ have been computed numerically for cylindrical samples with their axis of symmetry parallel to the applied field, [33-35]. Two types of averaging can be used to determine the effective value of $D$ [34]: (i) the "fluxmetric" average, which is an average over the cylinder cross-section located in the mid-plane of the sample, and (ii) the "magnetometric" average, which is an average over the entire cylinder volume. The relevant values of $D$ in the present study were obtained by a magnetometric average, given that the sample (although itself relatively large) is smaller than the pick-up coil.

The signal across the secondary (pick-up) coils, $v$, is proportional to the frequency $f$ and amplitude $H_a$ of the applied magnetic field and to the volume $V_s$ and the "external" susceptibility $\chi_{ext}$ of the sample [3, 31], i.e.;

$$v = \frac{1}{\alpha} H_a \, f \, V_s \, \chi_{ext} \qquad (2)$$

where the parameter $\alpha$ denotes the calibration constant (units A m$^2$ V$^{-1}$ s$^{-1}$) of the susceptometer and has to be determined via an appropriate calibration procedure. The calibration can be performed by using either mutual inductance calculations [31] for the case of long, thin specimens, or experimentally, using standards of known susceptibility and/or demagnetization factors. It should be noted, however, that the calibration factor derived by either technique is valid strictly only for specimens of the same size and shape as the standard used [31]. The relatively large size of the samples to be investigated in the present study enabled a *direct experimental* calibration by replacing the cylindrical superconducting sample by a copper wound coil of exactly the same size and external geometry, as shown schematically in Fig. 2. This calibration coil is made of a single layer of $N$ turns of fine copper wire and is located precisely at the sample position. The calibration coil is connected electrically in series with the primary coil of the susceptometer. This ensures that the calibration AC magnetic moment $m$ (produced by the calibration coil) has exactly the same frequency and the same phase as the applied AC magnetic field $H_a$ (produced by the primary coil). If $I$ represents the current in the calibration coil of cross-section $S$, the resulting magnetic moment, $m_{cal}$, is then given by;

$$m_{cal} = N \, I \, S \qquad (3)$$

The calibration coil is therefore equivalent to a magnetized sample of volume $V_s$ and whose volume magnetization, $M_{cal}$, is given by;

$$M_{cal} = \frac{m_{cal}}{V_s} = \frac{N \, I \, S}{S \, d} = \frac{N \, I}{d}, \qquad (4)$$

where $d$ is the length of the coil along its axis. In this case, the coil behaves as a sample of precisely known in-phase external susceptibility $\chi_{ext}$ given by;

$$\chi_{ext} = \frac{M_{cal}}{H_a}. \qquad (5)$$

The calibration process consists of measuring the voltage *v* across the pick-up coils in the presence of the "simulated" coil sample of susceptibility $\chi_{ext}$. In order to cancel the experimental error caused by a possible contribution of the offset signal to the measured signal, the calibration procedure can be performed twice : a first measurement is taken with the current in the calibration coil in one direction, a second measurement is taken with the current reversed in the calibration coil, and two voltages are substracted.

The knowledge of $\chi_{ext}$, *f*, $H_a$, $V_s$ and *v* enables the calibration constant $\alpha$ of the susceptometer (cf. Eq. (2)) to be determined for samples of the same volume and geometry as the calibration coil. As an example, the calibration constant for a cylindrical sample of thickness 12 mm and cross section 707 mm² was determined using a calibration coil of 64 turns of 150 µm diameter copper wire. The calibration constant was checked to be independent of both (i) the amplitude of the calibration current (which ranged from 40 µA to 4 mA), and (ii) the frequency of the calibration current (up to 1 kHz). The field uniformity was also checked during the calibration procedure; the signal was stable to within 1% for both radial and axial displacements of the calibration coil by up to 4 mm. This process produced a calibration constant $\alpha = 11.24$ A m² V⁻¹ s⁻¹, which is greater than equivalent values reported in the literature for high sensitivity AC susceptometers (e.g. $\alpha = 2.1546$ A m² V⁻¹ s⁻¹ in ref [3]). This leads to lower relative sensitivity for the current design. Nevertheless, equation (2) shows that the voltage *v* across the pick-up coils is proportional to the volume of the sample $V_s$, which, in the present case, is significantly larger than the volume accessible in classical susceptometer designs and compensates for the decrease in sensitivity.

*2.3. Heating system and cryogenic aspects*

The sample space in the susceptometer described here is connected to a vacuum rotary pump, enabling a medium vacuum (2 x 10⁻² mbar at room temperature) to be achieved. This vacuum is maintained during the whole experiment and no helium gas is consumed during the measurement. Four aluminized Plexiglas radiation shields were placed between the top cap of the sample chamber (at ambient temperature) and the position of the sample in order to mimize radiation heating. The presence of the thin aluminium layer deposited on the radiation shields was found to have no effect on the signal from the pick-up coil. The sample temperature is able to reach a temperature close to 77 K in approximately 90 minutes when the whole sample chamber arrangement (i.e. the part shown in Fig. 1(b)) is immersed in liquid nitrogen. The heater is then turned on, the temperature is increased slowly at a constant sweep rate (*dT / dt*) and AC susceptibility data are collected by a PC via a GPIB interface port. Several sweep rates are used during a single measurement run (e.g. 15 K /hour from 78 K to 88 K and 2.5 K/hour in the vicinity of the superconducting transition (88-95 K)) in order to minimize the total measurement time whilst allowing satisfactory temperature resolution.

It is difficult to achieve a relatively high heating rate (~15 K /hour) whilst maintaining a small thermal gradient within a large sample during the heating process. A bespoke heating arrangement was developed, therefore, in order to ensure sufficient temperature uniformity throughout the sample during the temperature sweep. This involved 'sandwiching' the sample between two heating plates located over and beneath its top and bottom faces. Each heating plate consisted of a circular Minco Thermofoil$^{TM}$ heater (R = 28.8 ohms at room temperature) adhered to a 1 mm thick alumina plate using Epoxy resin. Alumina is used for this application in view of its poor electrical and reasonably good thermal conductivity [37]. In addition, alumina is much cheaper than sapphire, which is used commonly as an electrically non-conducting heat sink. The use of two heating plates produces a significant reduction in thermal gradient across the sample. This was verified by numerical modeling using Quickfield® software. Fig. 3 shows the modeled temperature distribution within a sample heated either by (i) a single heating plate with an output power $P$ (Fig. 3(a)), or (ii) two heating plates, each of output power $P/2$ (Fig. 3(b)). The locations of the maximum and minimum temperatures are indicated by arrows in Fig. 3. It can be seen that these points are located along the lateral surface of the pellet (radius $R$) for both configurations, but the maximum temperature difference is approx. 4 times smaller when the two-heater arrangement is used. This system is therefore efficient in achieving simultaneously a fast heating rate and a small thermal gradient across the sample. Moreover, the two-heater configuration allows the system to reach a higher temperature due to reduced heat losses along the top and bottom faces of the sample.

In practice, the two heaters are connected in series and are fed with a constant current provided by a Keithley 2400 DC current source. In order to maintain a constant heating rate ($dT_{set}/dt$), the sample temperature ($T_{mes}$) is measured continuously and the heating power $P$ is adjusted using a simple proportional regulation control algorithm $P = P_0 + k\left(dT_{set}/dt - dT_{mes}/dt\right)$, where $P_0$ and $k$ are constants. The maximum temperature then depends simply on the maximum heater power. The heating power is limited to 640 mW and the maximum temperature is approximately 120 K in the present system.

It is essential to ensure that each heating plate produces a similar heat flux towards the sample if the thermal gradient within the sample is to be minimised. In order to achieve the highest possible thermal insulation between the sample and the sample chamber walls, the following method of supporting the sample was designed. The sample + heaters arrangement was placed in the bottom part of the chamber on a weak thermal link consisting of three small glass balls (2.4 mm in diameter) located at the vertices of a triangle. Glass and alumina each have relatively high Young's modulii, so that their effective contact cross-sectional area is relatively small [38], thereby producing in a very weak thermal link. As a consequence, the heat flux produced by the top and the bottom heating plates is directed mainly towards the sample, minimizing the thermal gradient, as required. In addition, an aluminized Mylar foil was adhered to the inner wall of the chamber to reduce heat loss by radiation. This foil was so placed that it makes an open circuit to prevent Eddy currents.

The sample temperature is measured using a Pt-100 platinum resistive temperature detector (RTD). The temperature sensor is insulated from the heat flux from the heating plates and placed in contact with the sample using Apiezon grease to ensure a good thermal link. The thermometer is connected to an Agilent 34401A multimeter by four, thin manganin wires (100 µm diameter) in order to minimize heat losses by conduction through the wires [39]. All devices (temperature measurement, current source and lock-in amplifier) are controlled using a GPIB USB-HS board connected to a personal computer running Matlab® 6.5 (see Fig. 4).

The maximum sampling rate is used throughout the measurement in order to refine the temperature step as far as possible. A computer program is used to predict the range of each successive measurement using a linear extrapolation over the five previous data points (a less satisfactory alternative is to use the auto-range mode of the lock-in amplifier). This procedure was found to be very efficient in generating accurate data for a minimum measurement time. In addition, as described above, a lower heating rate is set during the thermal approach to the critical temperature $T_c$. This enabled a temperature resolution around the critical temperature of about 4 mK to be achieved with a heating rate of 2.5 K/hour. This is ideal for an accurate measurement of the AC susceptibility signal in the vicinity of the superconducting transition.

## 3. Results and discussion

*3.1 Temperature uniformity measurement*

The temperatures at several locations on the lateral surface of a large bulk YBCO sample were recorded during the heating process using two differential thermocouples (shown in Fig. 5) in order to evaluate the performances of the heating system. Differential thermocouple #1 measures the temperature difference $\Delta T_1$ between two points located at 1.5 mm from the top and the bottom of the sample, whereas differential thermocouple #2 measures the temperature difference $\Delta T_2$ between one point located at 1.5 mm from the heating plate, and another at mid-height. Both thermocouples are made of fine (50 µm diameter) copper and constantan wires (type T). Silver paint was used to join the copper and constantan for each thermocouple, each anchored thermally over a length of 25 mm on the lateral surface of the sample. The joined regions are insulated electrically from the superconductor using thin cigarette paper [40]. The copper wires are twisted in order to avoid inductive pick-up and the top ends are thermalized together. The thermocouple voltages are read with a Keithley 2001 DC multimeter, enabling a resolution of 100 nV to be achieved.

A criterion imposed here is that the maximum temperature difference $\Delta T_{max}$ (measured either by $\Delta T_1$ or $\Delta T_2$) should always remain lower than 0.2 K during the sample heating process. Table 2 shows the value of $\Delta T_{max}$ for a given net heater input power of 100 mW for three 3 different cases: (i) 100 mW in the bottom heater only, (ii) 50 mW in each heater, and (iii) 100 mW in the top heater only. The three configurations lead to a similar average heating rate of the sample. The maximum temperature difference within the sample, however, is observed to be significantly lower when a 50 - 50 % sharing of the heat flux is achieved, rather than a single-plate heating arrangement, which is in agreement with the results of the thermal model shown in Fig. 3. As a result, it can be concluded that the thermal linkage between the heaters and the sample is very efficient and that the double heating plate system allows a very high heating rate to be achieved with excellent temperature uniformity within the sample.

*3.2 AC susceptibility measurement*

AC susceptibility measurements were performed on large, bulk superconducting (RE)BCO samples to illustrate the performance of the system. The materials were synthesised by employing a TMSTG process described elsewhere [22]. Two YBCO and SmBCO large grain superconductors processed in air (Samples A and B) were investigated as part of this study. Sample A was a large YBCO single domain pellet of diameter 30 mm and thickness 12 mm, whereas Sample B was a smaller, SmBCO pellet of diameter 18 mm and thickness 6 mm. The two samples differed in their microstructures, in addition to their different sizes: Sample A consisted of only one large single grain, whereas sample B contained one large grain at the centre of the pellet, one secondary (smaller) grain near the edge of the pellet and several small (~ micron-sized), homogeneous grains located in the remainder of the pellet, as shown schematically in the inset of Fig. 7. The typical sizes of the main and secondary grains in Sample B were 10 mm and 4 mm, respectively.

Results obtained for Samples A and B are shown in Figs. 6 and 7. The values of external susceptibility $\chi_{ext} = \chi'_{ext} - j\chi''_{ext}$ have been converted into equivalent internal values $\chi_{int} = \chi'_{int} - j\chi''_{int}$ using the classical formulas [3] derived from Eq. (1);

$$\chi'_{int} = \frac{\chi'_{ext} - D\left({\chi'_{ext}}^2 + {\chi''_{ext}}^2\right)}{D^2\left({\chi'_{ext}}^2 + {\chi''_{ext}}^2\right) - 2D\chi'_{ext} + 1}, \qquad (6)$$

$$\chi''_{int} = \frac{\chi''_{ext}}{D^2\left(\chi'^{\,2}_{ext} + \chi''^{\,2}_{ext}\right) - 2D\chi'_{ext} + 1}. \tag{7}$$

The demagnetization factors *D* for Samples A and B are the magnetometric values calculated by Chen et al. [34, 35] for cylinders of the same size as the two pellets, i.e. *D* = 0.5823 and 0.6016. It can be seen from Fig. 6 that the susceptibility measured at the lowest applied field (10 µT) for Sample A exhibits a plateau at low temperature, indicating that a state of low loss is achieved in this sample in this temperature range. The corresponding value of the in-phase internal susceptibility, $\chi'_{int}$, is − 0.963, which agrees well with the expected theoretical value for a perfect diamagnetic state ($\chi'_{int} = -1$). This agreement gives further confidence in the data recorded using this experimental system. The value of $\chi'_{int}$ at 78 K and 10µT for Sample B is − 0.967. This suggests a quasi-perfect diamagnetic state in this sample at low temperature, despite its relatively inhomogeneous microstructure.

The experimental system developed here facilitates the investigation of marked differences between the superconducting phase transitions of Samples A and B. A slow heating rate of 2.5 K/h in the vicinity of the transition temperature enables a large number of measurement points to be obtained in this region. As a result, the superconductor consisting of one large, single grain (Sample A) exhibits a critical temperature of ~ 91.3 K and a narrow superconducting transition, giving rise to a well-defined and sharp peak in $\chi''$, as can be seen from the inset in Fig. 6. In contrast, the $\chi''(T)$ curves for sample B, exhibit a shoulder (inset Fig. 7). The main peak occurring at $T \approx 91.7$ K for an applied field of 100 µT is accompanied by a distinct shoulder occurring around $T \approx 91.4$ K. This feature can be viewed as the convolution of two separate "peaks" that occur close to each other. $\chi''$ has a finite but small value and exhibits a flat bump at lower temperatures. The physical origin of this behavior can be understood briefly as follows. The AC magnetic properties of high-$T_c$ superconductors are known to be due to macroscopic shielding currents that circulate either within the grains (*intragranular* currents) or across the grain boundaries (*intergranular* currents) [41,42]. Previous experiments carried out on similar bulk melt-processed (RE)BCO materials have established that naturally-occurring grain boundaries (i.e. those forming during the melt growth process) are usually characterized by an intergranular current density that is much smaller than the intragranular value [43]. It is likely, therefore, that the "double peak" structure present in the inset of Fig. 7 is exclusively of an intragranular origin; the temperature separation between both "peaks" is thought to arise from the difference in relative size of the two large grains (i.e. the main grain and the sub-grain due to secondary nucleation). According to the Bean model [44], a peak in the $\chi''(T)$ curve corresponds to complete penetration of supercurrents to the centre of their circular paths [45], i.e.;

$$H_a = J_c(T_{peak}) \cdot R, \tag{8}$$

where $H_a$ is the applied magnetic field, $J_c(T_{peak})$ the critical current density at the χ" peak temperature and $R$ is the characteristic radius of the paths of the induced shielding current. Therefore, the magnetic shielding occurring in the largest grain (average diameter ~ 10 mm) will result in a χ" peak corresponding to a small current density, i.e. at a high temperature. In the case of the smaller, secondary grain (average diameter ~ 4 mm), the intragranular peak is expected to correspond to a slightly larger current density, i.e. at a slightly lower temperature. The behaviour of the AC magnetic signal at low temperature ($T < 89$ K) is likely to be due to the contribution of the very small grains (of micron size) present in the remainer of the pellet.

In summary, the data presented in Figs. 6 and 7 show, for the first time, the temperature dependence of the AC magnetic properties measured on *whole,* large bulk HTS samples. It should be emphasized that the fine structure arising in the χ" peak of sample B (Fig. 7) is due precisely to shielding currents circulating on different macroscopic length scales of the order of ~ 10 mm. These features would not appear if the sample was cut into smaller specimens that were subsequently characterized individually in a classical AC susceptometer. These experimental results also underline the critical importance of a fine temperature resolution of the experimental system in order to resolve peaks that occur at similar temperatures (i.e. within ~ 0.3 K). Finally, it should be noted that the temperature uniformity check described in section 3.1 has demonstrated clearly that the temperature gradient appearing within the sample is smaller than 0.1 K when a heating rate of 25 K/h is employed. The temperature gradient in the vicinity of the critical temperature is expected to be significantly smaller than 0.1 K due to the relatively low heating rate (2.5 K/h). This gives confidence that the double-peak structure is not an experimental artifact caused by an inhomogeneous temperature distribution within the sample under study. The experimental arrangement described in this paper, therefore, provides an accurate and reliable technique for the AC magnetic characterization of large, bulk superconducting samples.

## 4. Conclusions

A simple and reliable susceptometer working in the temperature range 78-120 K and capable of measuring the AC susceptibility of large, bulk superconducting samples up to 32 mm diameter has been designed and constructed. The ability to characterize such large samples is not possible in conventional susceptometers or magnetometers. The system incorporates a bespoke heating device consisting of two heating plates located above and beneath the top and bottom faces of the sample to enable fast and precise measurements in the "temperature sweep" mode. As a result, a relatively fast heating rate (25 K/hour) can be achieved whilst a relatively small temperature gradient (< 0.1 K) is maintained over the sample. An AC susceptibility measurement between 78 K to 95 K takes typically approximately 2 hours with the new system, with a 4 mK temperature resolution of around $T_c$. The apparatus can be calibrated precisely using a copper coil connected electrically in series with the primary coil. The susceptometer was used to measure the temperature dependence of the AC susceptibility of large, bulk melt-processed YBCO and SmBCO samples. Although the apparatus has been optimized for the study of the phase transition of superconducting materials, it could be applied relatively easily to other large magnetic samples that would need to be characterized in an AC magnetic field, and possibly in the presence of a superimposed DC magnetic field.

**Acknowledgements**

Ph. L. is grateful to the "Communauté Française de Belgique" for a travel grant. We acknowledge K. Iida, Y. H. Shi and T. D. Withnell for fruitful discussions and comments. The technical help of M. Hansez, P. Harmeling, J. Simon has been greatly appreciated. We also thank the FNRS, the ULg, the UK EPSRC and the Royal Military Academy (RMA) of Belgium for cryofluid and equipment grants.

**Tables**

**Table 1**. Susceptometer coil geometries. Note that the experimental system contains 2 quasi-identical primary coils and 2 quasi-identical secondary coils (cf. figure 1).

|                  | **Primary coil** | **Secondary coil** |
|------------------|------------------|--------------------|
| Length           | 120 mm           | 20 mm              |
| Outer diameter   | 50 mm            | 40 mm              |
| Wire diameter    | 0.6 mm           | 0.15 mm            |
| Number of turns  | 2800             | 400                |
| Number of layers | 14               | 2                  |

**Table 2**. Maximum temperature difference for 3 heating configurations corresponding to a total input power $P_{tot}$ of 100 mW

| Bottom heater ($P_{Bottom}/P_{tot}$) | Top heater ($P_{Top}/P_{tot}$) | $\Delta T_{max}$ (K) | Heating rate (K/h) |
|---|---|---|---|
| 100 % | 0 %   | ≈ 0.25 | ≈ 25 |
| 50 %  | 50 %  | < 0.1  | ≈ 25 |
| 0 %   | 100 % | ≈ 0.25 | ≈ 25 |

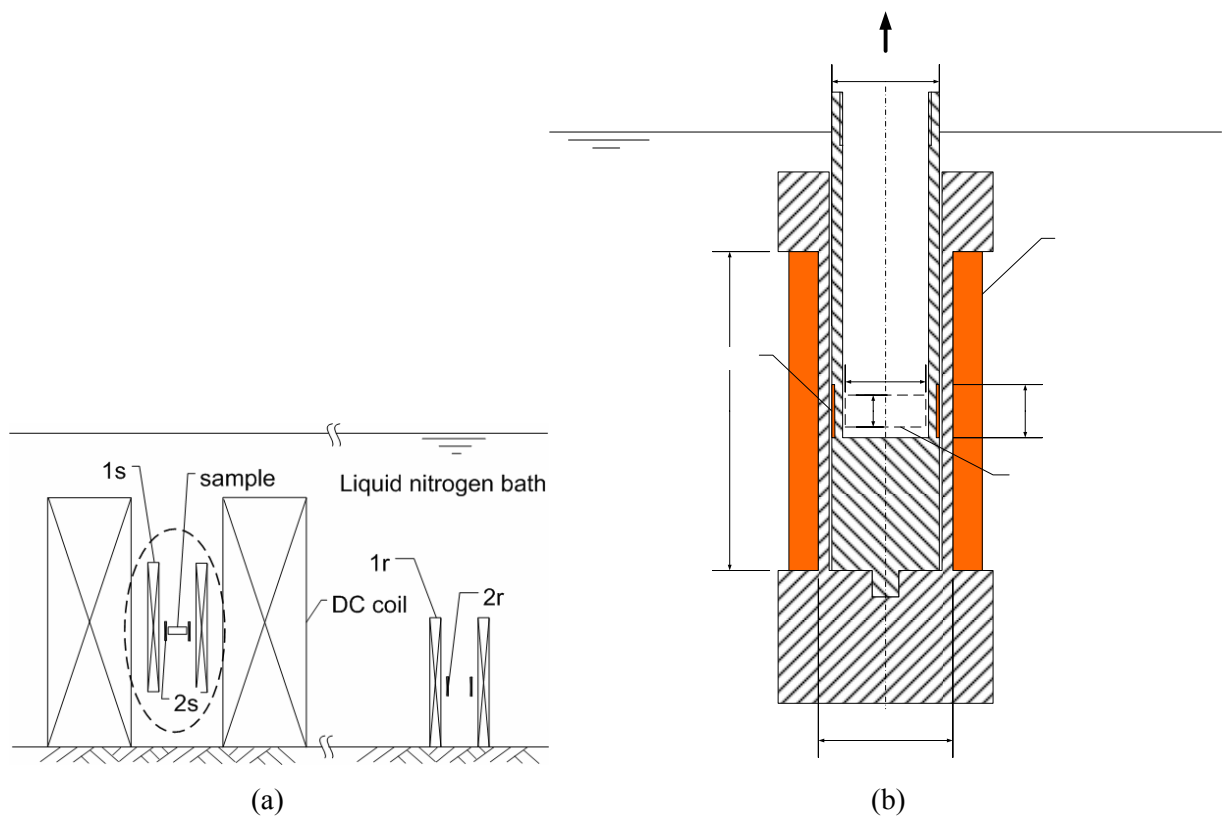

**Figure 1**. (a) Schematic illustration of the geometrical arrangement of the coils used for the AC susceptometer. The primary coil consists of two separate coils, one containing the sample (coil "1*s*") and one used for reference (coil "1*r*"). Coil "1*s*" can be inserted in a large coil that can generate a DC magnetic field. Primary coils "1*s*" and "1*r*" contain their own secondary coil, "2*s*" and "2*r*", respectively. (b) Cross-section of the part of the susceptometer containing the sample illustrating the geometry the primary coil "1*s*" and the secondary coil "2*s*". All dimensions are given in mm. The sample, located in a vacuum vessel, is centered in the secondary coil "2*s*". The reference coils ("1*r*" and "2*r*") are identical to those shown in Fig. 1b.

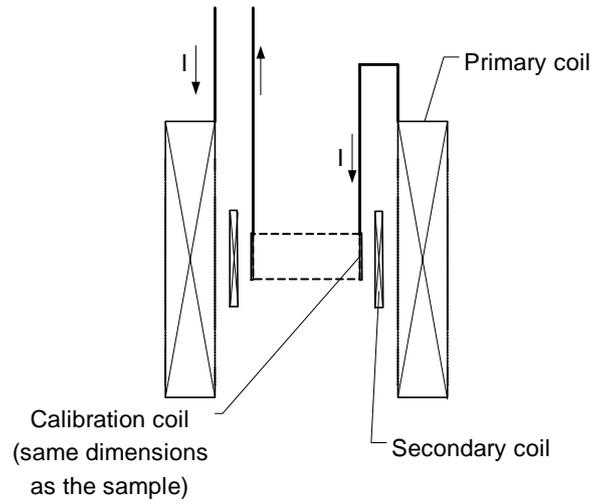

**Figure 2**. Schematic illustration of the method used to calibrate the susceptometer. A copper-wound calibration coil of the same dimensions as the sample to be studied is fed with the same current as the primary coil. This calibration coil is centered in the pick-up coil.

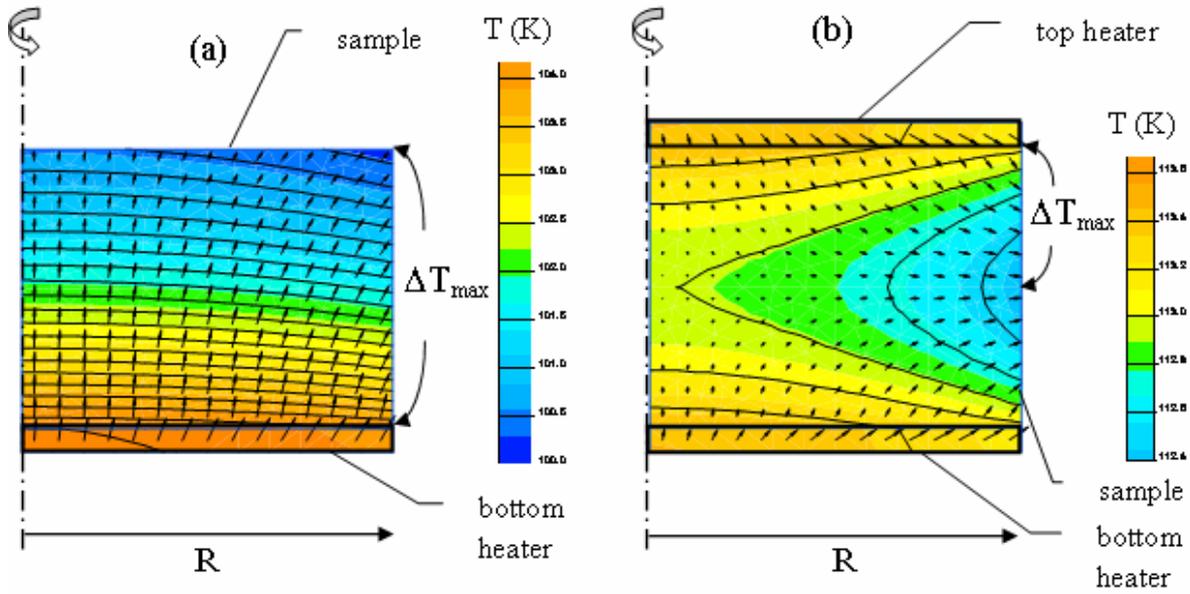

**Figure 3**. Modeled temperature distribution within a disc-shaped sample (radius *R*) heated either by (a) one or (b) two heaters. A surface heat flux of 1000 W/m² or 500 W/m² is imposed on the external surface of the alumina plate using either a (a) one heater or (b) two-heater arrangement, so that the total power output is the same in both cases. A convection boundary condition (convective heat transfer coefficient h = 15 W/m²K) is imposed on the other surfaces. The thermal conductivity values of the sample are those of YBCO, i.e. 7.5 and 2 W m$^{-1}$ K$^{-1}$ parallel to the radial and axial directions, respectively [36] and 30 W m$^{-1}$ K$^{-1}$ for the alumina plates [37].

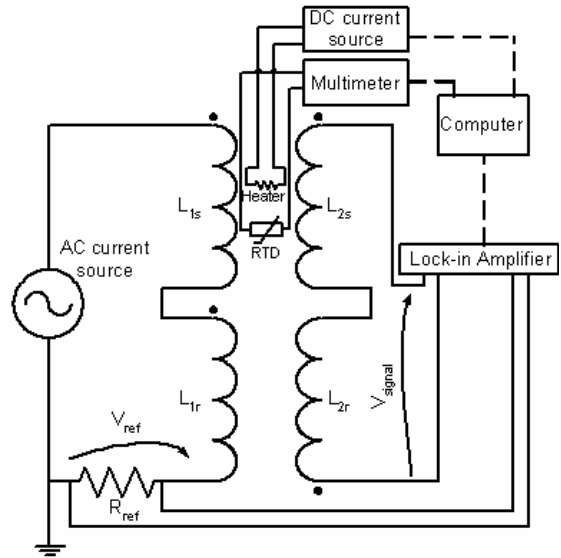

**Figure 4**. Schematic illustration of the circuit diagram. The GPIB connections are indicated by dashed lines. The symbols $L_{1s}$ and $L_{1r}$ refer to the primary coils with and without the sample present, respectively, whereas $L_{2s}$ and $L_{2r}$ refer to the corresponding secondary coils.

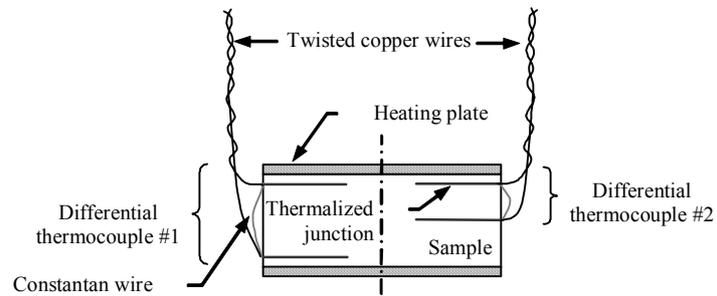

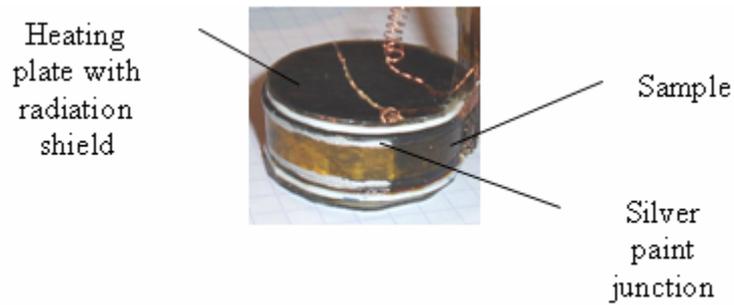

**Figure 5**. Top: schematic representation of the location of the two differential thermocouples used to determine the temperature distribution during heating of a 30 mm diameter, superconducting sample under conditions of constant power. Both thermocouples are placed on the lateral surface of the sample. Bottom: photograph of the fully instrumented sample.

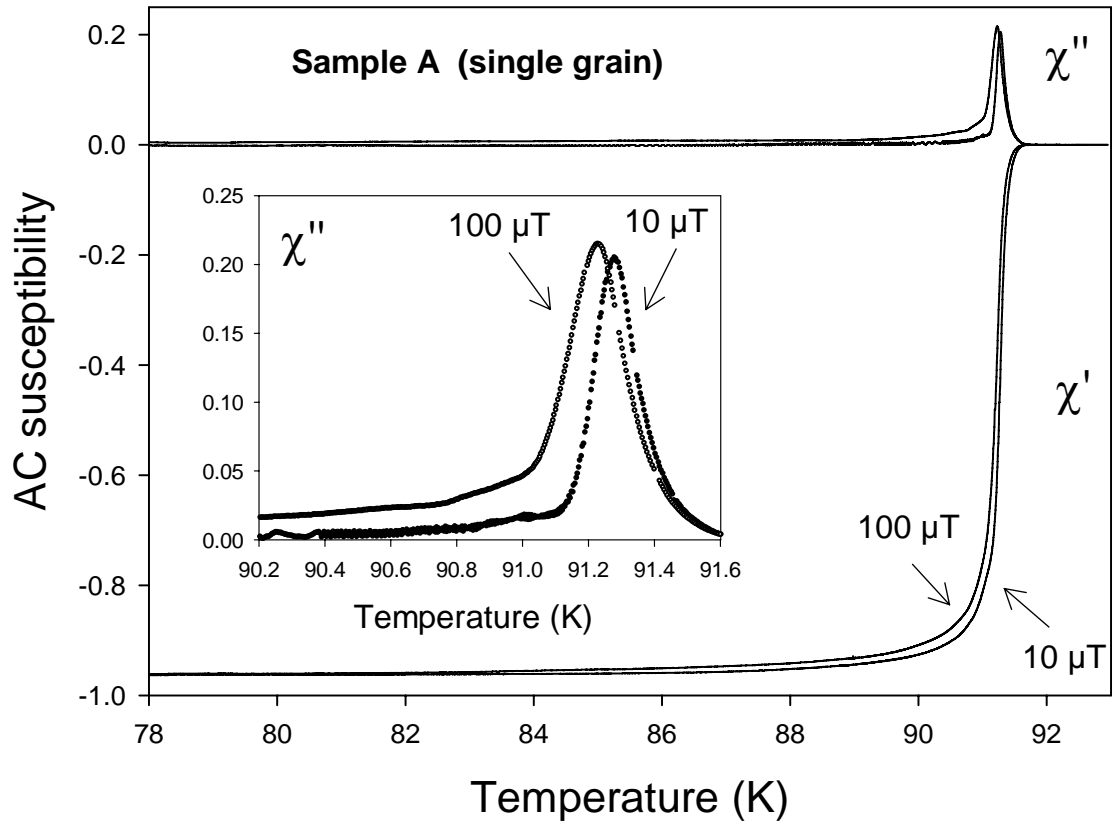

**Figure 6**.  Real ($\chi'$) and imaginary ($\chi''$) parts of the internal AC susceptibility vs. temperature measured for a large, single grain YBCO pellet (Sample A). A fixed frequency $f$ = 103 Hz was used and the amplitude of the AC field was either 10 μT or 100 μT. The inset shows an enlargement of the $\chi''$ peak.

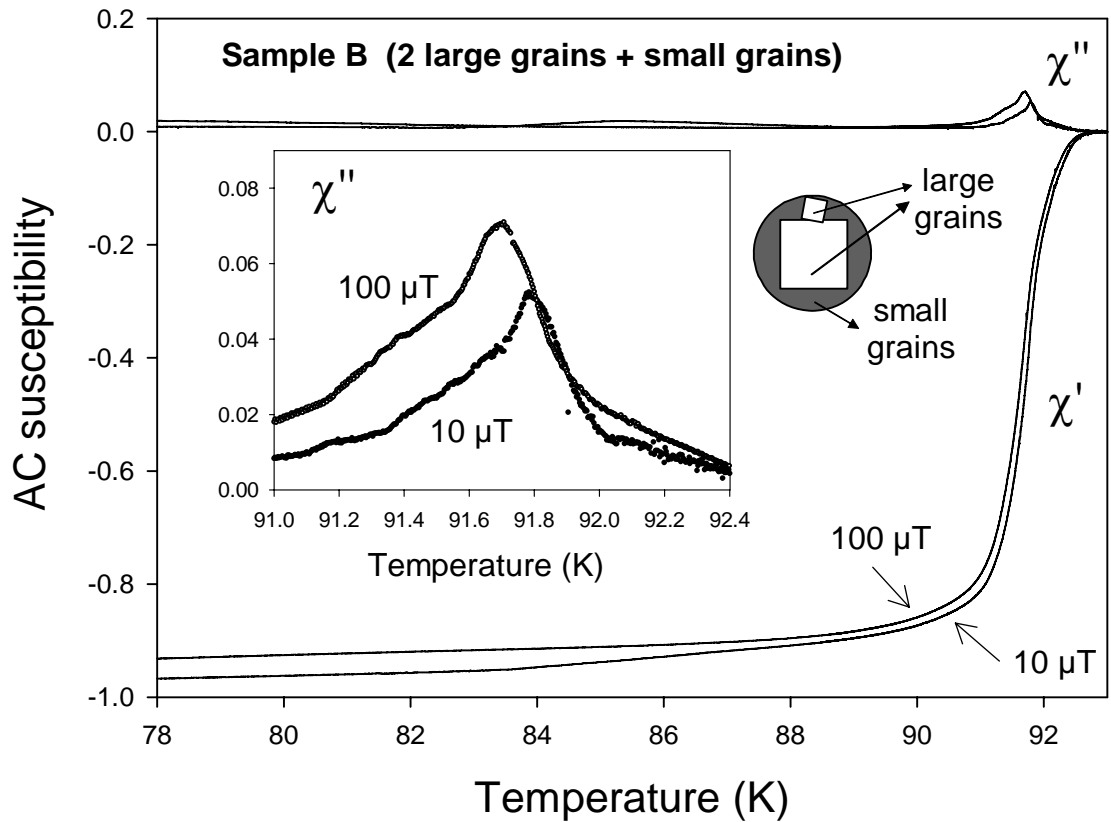

**Figure 7**. Real ($\chi'$) and imaginary ($\chi''$) parts of the internal AC susceptibility vs. temperature measured on a large SmBCO pellet (Sample B) containing 2 large grains of different sizes and a large number of small grains, as shown schematically in the in the inset. A fixed frequency $f = 103$ Hz was used and the AC field amplitude was either 10 µT or 100 µT. The inset shows an enlargement of the $\chi''$ peak.